\documentclass[aps,prl,twocolumn,showpacs,superscriptaddress,floatfix]{revtex4}
\usepackage{graphicx}
\usepackage{amsfonts}
\usepackage[figuresright]{rotating}  
\usepackage{amssymb}
\usepackage{amsmath}
\usepackage{subfigure}
\usepackage{multirow}
\usepackage{tabularx}

\def\C60{A$_x$C$_{60}$}

\def\HgCu3{HgCa$_2$Cu$_3$O$_{8+y}$}
\def\HgCu4{HgBa$_2$Ca$_3$Cu$_4$O$_{10+y}$}
\def\TlCu{Tl$_2$Ba$_2$CuO$_{6+\delta}$}
\def\TlCu3{Tl$_2$Ba$_2$Ca$_2$Cu$_3$O$_{10+y}$}
\def\TlCu4{Tl$_2$Ba$_2$Ca$_3$Cu$_4$O$_{12+y}$}

\def\BiCu3{Bi$_2$Sr$_2$Ca$_{2}$Cu$_3$O$_y$}

\def\8LSCO{La$_{1.88}$Sr$_{.12}$CuO$_4$}
\def\110LNSCO{La$_{1.5}$Nd$_{0.4}$Sr$_{0.1}$CuO$_{4}$}
\def\stage4LCO{La$_{2}$CuO$_{4+\delta}$}
\def\Y248{YBa$_2$Cu$_4$O$_8$}

\def\NbSe2{NbSe$_2$}
\def\TaSe2{TaSe$_2$}
\def\TiSe2{TiSe$_2$}
\def\NaCoOH2O{Na$_{0.3}$CoO$_{2y}$H$_2$O}
\def\MgB2{MgB${}_2$}

\def\avg#1{\langle#1\rangle}
\def\Re {\mbox{Re}}
\def\Im {\mbox{Im}}

\begin{document}
\title{Topological Insulators and Nematic Phases
from Spontaneous Symmetry Breaking in 2D Fermi Systems with a Quadratic Band Crossing}
\author{Kai Sun}
\affiliation{Department of Physics, University of Illinois at Urbana-Champaign, 
1110 West Green Street, Urbana, Illinois 61801-3080, USA}
\author{Hong Yao}
\affiliation{Department of Physics, Stanford University, Stanford, California 94305, USA}
\author{Eduardo Fradkin}
\affiliation{Department of Physics, University of Illinois at Urbana-Champaign, 
1110 West Green Street, Urbana, Illinois 61801-3080, USA}
\author{Steven A. Kivelson}
\affiliation{Department of Physics, Stanford University, Stanford, California 94305, USA}

\begin{abstract}
We investigate the stability of a quadratic band-crossing point (QBCP) in 2D fermionic systems. 
At the non-interacting level, we show that a QBCP exists and is topologically stable 
for a Berry flux $\pm 2\pi$, if the point symmetry group has either 
fourfold or sixfold rotational symmetries. 
This putative topologically stable free-fermion QBCP is marginally unstable 
to {\em arbitrarily weak}
short-range repulsive interactions. We consider 
both spinless and spin-$1/2$ fermions.
Four possible ordered states result: 
a quantum anomalous Hall phase, a quantum spin Hall phase, a nematic phase,  and a
nematic-spin-nematic phase.
\end{abstract}
\date{\today}
\maketitle 

{\it Introduction}---
In multi-band fermionic systems, a band-crossing point (BCP)  is a point
in the 
Brillouin zone where two 
bands cross. 
As the chemical potential reaches a BCP, the Fermi surface shrinks to a point
and new phenomena, not described by a Fermi liquid, result. 
The simplest and
best studied is the case of a linear band crossing, whose low energy physics is described by 
a Dirac fermion.  
Dirac fermions 
are a good description of the low energy states of 
 nodal superconductors, graphene and zero gap semiconductors.

In general, 
a Dirac point in a band structure is robust with respect to small changes in the effective potential which 
preserve the 
symmetries of the crystal, as has been extensively shown in various contexts~\cite{nielsen-1981-I,Kogut-1983,Haldane1988,Hatsugai-2006}. 
Moreover, short-range electron-electron interactions are perturbatively irrelevant for space dimension $d>1$.
Thus, there is a stable phase with free gapless Dirac fermions which becomes unstable 
above a critical interaction strength corresponding to a quantum critical point  beyond which lie 
phases with spontaneously broken space 
or point group symmetries and/or broken time-reversal invariance~\cite{wilson-1973,vojta-2000,Raghu2008}.

In this letter we consider a system with a quadratic band crossing point (QBCP) somewhere 
in its 2D Brillouin zone. This problem has not been discussed in depth, and only a few aspects have been analyzed.
The perturbative stability of a QBCP was studied for 2D
noninteracting systems with $C_{4v}$ symmetry in Ref. \cite{Chong2008}. 
For interacting fermions, it was noted in Ref. \cite{Sun2008} that a QBCP in 2D has 
instabilities, for arbitrarily weak interactions,  leading to the spontaneous breaking of rotational 
symmetry (nematic phase) or  time-reversal invariance, but its consequences were not explored in depth.

We begin by analyzing the general symmetry principles that protect a QBCP in lattice models
of noninteracting fermions. We find that QBCPs are protected  by time-reversal symmetry
 and $C_4$ or $C_6$ rotational symmetry. Explicit examples of lattice models 
 with both symmetries are presented. 
We 
show that 
short-range repulsive interactions are marginally relevant in the renormalization group (RG) sense. 
The symmetry breaking phases and phase transitions, 
both quantum and thermal, are  investigated by an RG analysis
and mean-field approximations, both presumably reliable at weak-coupling.

We determined the structure of the phase diagrams for both spinless and spin-$1/2$ fermions. 
In the spinless case the leading weak-coupling instability is to a gapped phase with broken time-reversal 
invariance, 
a quantum anomalous Hall (QAH) effect, and topologically protected edge states. 
For stronger interactions, there is
 a subsequent transition to a nematic  Dirac  
phase and an intermediate phase with QAH-nematic coexistence. 
For spin-$1/2$ fermions the phase diagram is more complex:
in addition to spin singlet QAH and nematic phases, there are also a spin triplet 
quantum spin Hall (QSH) phase\cite{comment} and a nematic-spin-nematic (NSN) phase \cite{Oganesyan2001,Kivelson2003-short,Wu2007a}.

{\it Quadratic band-crossing point}---
A BCP carries quantized Berry flux \cite{Haldane2004} as required by time-reversal symmetry:
$-i \oint_{\Gamma} d\mathbf{k} \cdot \avg{\psi(\mathbf{k})|\boldsymbol{\nabla}_{\mathbf{k}}|\psi(\mathbf{k})}= 
n \pi$,
where $\Gamma$ is a contour in the momentum space enclosing the BCP, $\psi(\mathbf{k})$ is the Bloch 
wave function 
in a band involved in the band crossing, and $n$ is an integer. For a Dirac point, the Berry flux is $\pm \pi$. 
Instead, the Berry flux at a QBCP is either $0$ or $\pm 2 \pi$ \cite{Sun2008}. The zero flux QBCP is an accidental 
band crossing, which can be removed by infinitesimal band mixing without breaking any symmetries, but a QBCP with 
$\pm 2\pi$ flux is robust and more interesting. 

A natural question to ask is if a QBCP is protected by the symmetries of the non-interacting system. 
In general, there are two ways to 
remove a QBCP. One way is to split it into several Dirac points while preserving the total Berry flux.  
A QBCP with flux $2\pi$, for example, can be split into two separate  Dirac points each with flux $\pi$, 
or three 
Dirac points with flux 
$\pi$ and one additional Dirac point with $-\pi$. The former case 
in general breaks the point group symmetry leaving, at most, 
a two-fold rotational symmetry unbroken. The latter case can take place while preserving a three-fold 
rotational 
symmetry, such as the case of bilayer graphene \cite{CastroNeto2009-short,Nilsson2008-short}. For a QBCP with
a fourfold or sixfold symmetry axis, the split into Dirac points 
 cannot occur without breaking that symmetry.
The alternative is to open a gap by breaking time-reversal symmetry 
or a symmetry that is formally  similar, such as the combined space and spin symmetry whose breaking leads 
to a spontaneous quantum spin Hall state, as discussed below. 
Thus, for a QBCP 
(with Berry flux $\pm 2\pi$) to be stable without fine tuning, two conditions are required: 
a)  the system must be time-reversal invariant 
and b) the QBCP must have $C_4$ or $C_6$ symmetry. 

\begin{figure}[hbt]
\begin{center}
\subfigure[]{\includegraphics[width=0.2\textwidth]{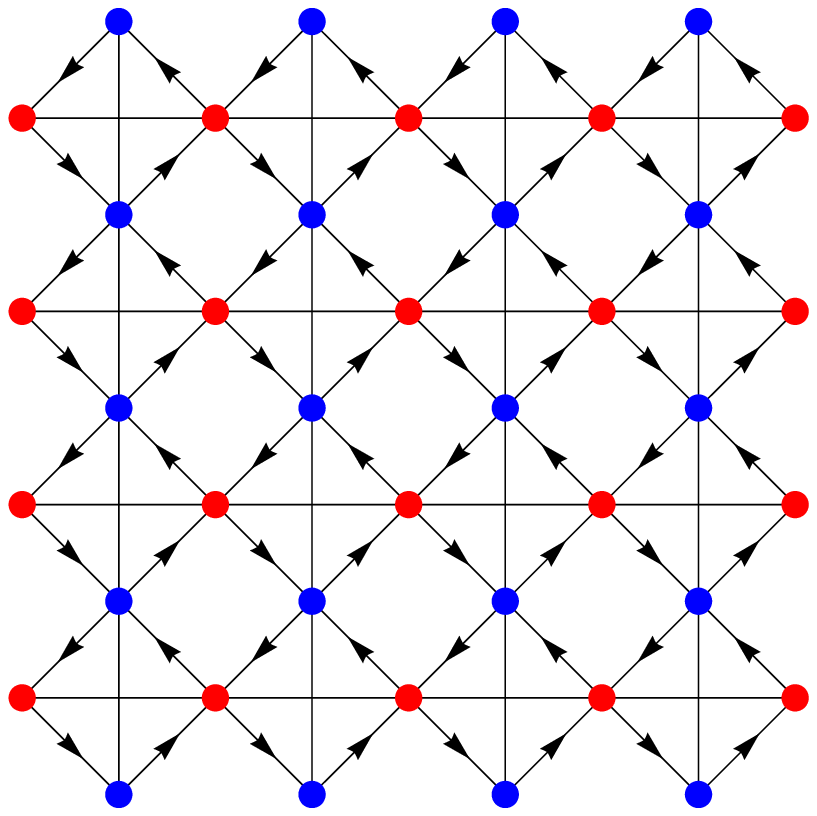}\label{fig:checkerboard}}
\subfigure[]{\includegraphics[width=0.2\textwidth]{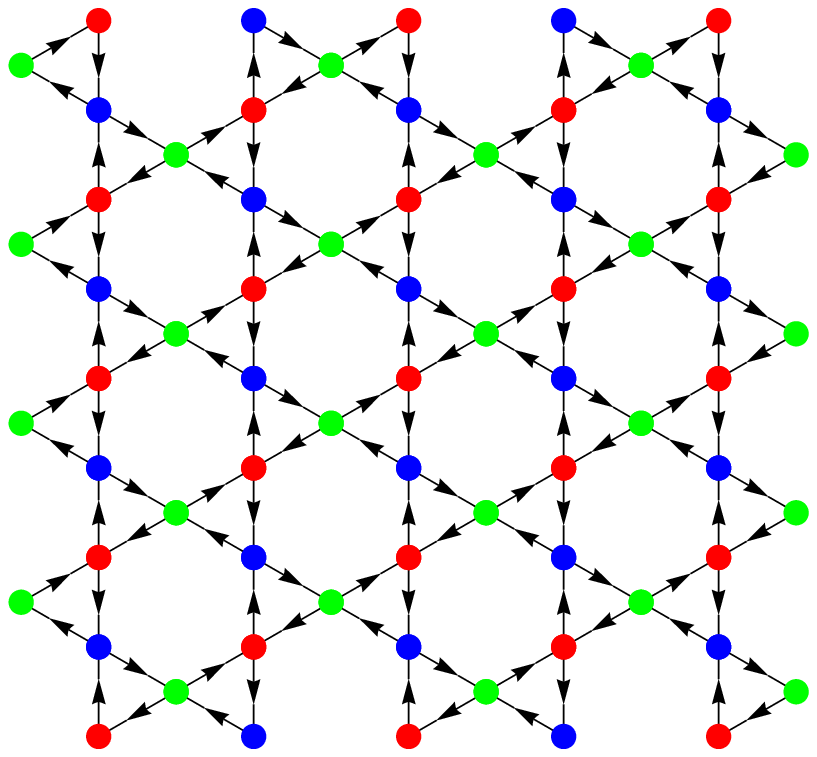}\label{fig:kagome}}
\end{center}
\caption{(Color online) (a) A checkerboard lattice and (b) a Kagome lattice. The
arrows represent currents in a spontaneously generated QAH state that 
breaks the time-reversal symmetry. See text for details.}
\end{figure}

An example of a QBCP in 2D with $C_4$ symmetry can be found in the checkerboard lattice \cite{Sun2008} 
Fig. \ref{fig:checkerboard}. This lattice can be regarded as the 2D projection of a 3D pyrochlore 
lattice. It is also the oxygen lattice in a CuO$_2$ plane of the cuprates. 
With one orbital per site, there are two bands 
crossing at a QBCP at $(\pi,\pi)$ 
with a fourfold rotational 
symmetry. 
At half filling the QBCP is at the Fermi level.
An example of a QBCP with $C_6$ symmetry 
is  a tight-binding model on a Kagome lattice, 
which has three bands. The middle band touches the bottom band at $(0,0)$, resulting in a 
QBCP with sixfold rotational symmetry. 
It lies at the Fermi level at  $1/3$  filling. 

In the presence of weak interactions, a BCP may become unstable if interactions are relevant 
in the RG sense. In 2D, a QBCP has a finite one-particle DOS, which implies that short-range interactions 
are marginal at tree level. 
We will show below, that at a 2D QBCP a short-range repulsive interaction is marginally relevant, 
and destabilizes this free fermion fixed point in weak coupling, leading to a state which spontaneously 
breaks one of the symmetries that otherwise would protect the QBCP.

{\it General model}---
We first formulate the theory of possible symmetry breaking 
phases in a general way. We begin with the spinless-fermion
case. Near a QBCP, in the low energy regime we have two species of interacting charged Fermi fields, 
$\psi_1$ and 
$\psi_2$, whose Hamiltonian is
\begin{align}
\!\!\!H\!=\!\int\!\! d\mathbf{r} \left[
\boldsymbol{\Psi}^\dagger(\mathbf{r})
\mathcal{H}_0
\boldsymbol{\Psi}(\mathbf{r})
+V \psi^\dagger_1(\mathbf{r})\psi_1 (\mathbf{r})\psi^\dagger_2 (\mathbf{r})\psi_2 (\mathbf{r})\right],
\label{eq:Hamiltonian}
\end{align}
where 
$\boldsymbol{\Psi}^\dagger=(\psi^\dagger_{1}, \psi^\dagger_{2})$, $\boldsymbol{\Psi}$ is its conjugate,
and $V$ is the coupling constant of the interaction.

The band structure near the QBCP is obtained by diagonalizing $2 \times 2$ Hermitian matrix 
$\mathcal{H}_0(\mathbf{k})$
for all Bloch-wave vectors in the neighborhood of the band crossing point, $|\mathbf{k}| \ll 1$.  
Quite generally we can choose the identity matrix 
$I$ and  the two real Pauli matrices $\sigma^x$ and $\sigma^z$ as a basis \cite{footnote-two_Pauli} and write 
$\mathcal{H}_0(\mathbf{k})$ as \cite{Blount1962}
\begin{align}
\mathcal{H}_0(\mathbf{k})=d_I I+d_x \sigma^x+d_z \sigma^z,
\label{eq:kinetic_Ham}
\end{align}
where $d_I =t_I  (k_x^2+k_y^2)$, $d_x =2 t_x k_x k_y$, and $d_z =t_z (k_x^2-k_y^2)$.
The $d$-wave symmetry of $d_x$ and $d_z$ distinguishes a QBCP from a Dirac point in which their counterparts 
have a $p$-wave symmetry. It is this $d$-wave nature that gives rise to the $\pm 2\pi$ Berry phase of 
a QBCP. For a QBCP with a $C_6$ rotational symmetry, $|t_x|=|t_z|$. If the system has particle-hole symmetry, 
$t_I=0$.
The condition $|t_I|< |t_x|$ and $|t_I|< |t_z|$ is required to ensure that away from the QBCP, one of the bands
lies above the degenerate point and the other band lies below.

At $V=0$, in the model of Eq. \eqref{eq:Hamiltonian}, the fermions have a finite DOS but do not have a Fermi 
surface. They have a dynamic critical exponent $z=2$, and an effective dimension 
$d_{\rm eff}=d+z=4$ \cite{Sachdev1999}. $\boldsymbol{\Psi}$ has dimension one $[\boldsymbol{\Psi}]=1$, and 
the only local four-fermion operator allowed is
marginal 
 since  $4[\boldsymbol{\Psi}]=d_{\rm eff}$. 
There is a single 
dimensionless coupling constant $g=V/|t_x|$. 

This system is similar to $d=1$ spinless fermions, a system with two Fermi points and 
dynamic critical exponent $z=1$. In the 1D case the Fermi field has scaling dimension $1/2$,
so there is only one interaction, four-Fermi backscattering, which is
potentially important. However, due to a cancellation between the Cooper channel and 
the bubble term in 1D, the interaction is exactly marginal to all orders in 
perturbation theory\cite{emery79}, which is the origin of Luttinger 
liquid behavior in 1D. In contrast, no similar cancellation occurs for fermions in 2D with 
$z=2$. Although the 4-Fermi interaction is superficially marginal it is actually marginally relevant. 
We find, that to one-loop order, the RG beta function for $g=V/|t_x|$ is 
\begin{equation}
\beta(g)=\frac{d g}{d l}= \alpha
g^2+ O(g^3),
\label{eq:beta}
\end{equation}
 where $\alpha=\frac{1}{2\pi^2}K \left(\sqrt{1-(t_z/t_x)^2}\right)$, 
 $l$ is a momentum rescaling $k\rightarrow k e^{-l}$, and $K(x)$ is the complete elliptic integral.
For $|t_x|=|t_z|$, {\it i.e.\/} a QBCP with $C_6$
symmetry, 
$\alpha=(4\pi)^{-1}$ \cite{band-anisotropy}.
Hence, Eq.\eqref{eq:beta} implies that for $g>0$ the effective coupling constant flows to strong coupling. 

To explore the consequences of this instability, we investigated,
in a mean-field level, possible orderings of bilinear order parameters: 
\begin{eqnarray}
&& \Phi=\avg{\boldsymbol{\Psi}^\dagger(\mathbf{r})\sigma_y \boldsymbol{\Psi}(\mathbf{r})}, 
\\
\label{eq:order_parameters}
&&Q_1=\avg{\boldsymbol{\Psi}^\dagger(\mathbf{r}) \sigma_z \boldsymbol{\Psi}(\mathbf{r})},~
Q_2=\avg{\boldsymbol{\Psi}^\dagger(\mathbf{r}) \sigma_x \boldsymbol{\Psi}(\mathbf{r})}.
\nonumber
\end{eqnarray}
$\Phi$ is the order parameter
of a time-reversal symmetry breaking gapped QAH  phase \cite{Haldane1988,Raghu2008}. This phase has a zero-field 
quantized Hall conductivity $\sigma_{xy}=e^2/h$. 
$Q_1$ and $Q_2$ describe the nematic phases in which the 
$C_4$ or 
$C_6$ rotational symmetry is broken down to $C_2$ by splitting the QBCP into two Dirac points located 
along the direction of one of the main axes
($Q_1$), or along a diagonal ($Q_2$).
The nematic phase is an anisotropic semimetal. Unlike in graphene, 
where the two Dirac points have  Berry fluxes $\pi$ and $-\pi$,
in the nematic phase both Dirac points  have the same 
Berry flux. 
There is also a phase in which nematic  ($Q_{1}\neq 0$ or $Q_{2}\neq 0$) 
and QAH orders
($\Phi \neq 0$) coexist, an insulating  analog of the metallic time-reversal breaking nematic $\beta$ 
phases of Ref. \cite{Sun2008}.

Since there is only one coupling constant ($V$) in Eq.\eqref{eq:Hamiltonian}, the  weak-coupling ordering 
tendencies are determined by the logarithmically divergent normal state
susceptibilities $\chi_\Phi$ 
(QAH order) and  $\chi_{Q_1}$ and $\chi_{Q_2}$ (nematic order). For general $t_x$ and $t_z$, they
satisfy  $\chi_\Phi=\chi_{Q_{1}}+\chi_{Q_{2}}$. 
Hence,  $\chi_\Phi> \chi_{Q_i}$ ($i=1,2$), so the leading weak coupling instability is to the (gapped)  QAH state.

The mean-field Hamiltonian is
\begin{eqnarray}
&&\!\!\!\!\!\!\!\!\! H_{\rm MF}=\int d\mathbf{r} \Psi^\dagger(\mathbf{r}) 
\left[\mathcal{H}_0-\frac{V}{2}\left(Q_1 \sigma_z + Q_2 \sigma_x + \Phi \sigma_y\right)\right] 
\Psi(\mathbf{r})\nonumber \\
&&+\frac{V}{4} \int d\mathbf{r} \left(Q_1^2+Q_2^2+\Phi^2\right)
\label{eq:MF}
\end{eqnarray}
By minimizing the ground state energy of $H_{\rm MF}$ we find that at weak coupling the ground state is 
indeed the QAH phase, with a gap $\Delta \sim \Lambda  \exp(-2 /\alpha g)$ ($\Lambda$ is a cutoff) and a mean-field critical temperature 
$T_c \sim \Delta$, consistent with the scaling predicted by the RG. 
A 3D example of the QAH at finite coupling is discussed in Ref. \cite{Zhang2009}.

Mean-field theory also predicts nematic phases provided that irrelevant operators, 
such as $\int d \mathbf{r}d \mathbf{r'} \sum_{i=1,2} 
U(\mathbf{r}-\mathbf{r}')\psi^\dagger_i (\mathbf{r}) \psi_i (\mathbf{r})\psi^\dagger_i (\mathbf{r}')
\psi_i (\mathbf{r}')$, are also included. 
The nematic phase $Q_1$ is energetically favored at small $V>0$ and $U<0$ if $|U/V|$ is large enough. 
As $|U/V|$ is reduced, the nematic phase gives way to the QAH phase (and to a mixed phase).

{\it Lattice models}--- 
We consider the following minimum model on a checkerboard lattice with a QBCP. 
\begin{equation}
	H=\sum_{ij}-t_{ij}c^\dag_i c_j +V \sum_{\langle ij\rangle} c^\dag_i c_i c^\dag_j c_j,
	\label{eq:lattice_model}
\end{equation}
where $t_{ij}$ is the hopping amplitude between sites $i$ and $j$ and $V>0$ is the nearest-neighbor repulsion. 
Here, $t_{ij}=t, t', t''$, respectively  for nearest neighbors, and next-nearest neighbors 
connected (or not) by a diagonal bond, Fig. \ref{fig:checkerboard}. There are two sublattices 
$A$ (red) 
and $B$ (blue). The fermion spinor is $\Psi^\dag =(c^\dag_{A},c^\dag_{B})$. 
The parameters of the free fermion Hamiltonian [Eq. \eqref{eq:kinetic_Ham}] 
are  $d_I=-(t'+t'')(\cos k_x+\cos k_y)$, $d_x=-4t\cos \frac{k_x}{2}\cos\frac{k_y}{2}$, 
and $d_z=-(t'-t'')(\cos k_y -\cos k_x)$. The QBCP is  $M=(\pi,\pi)$, at the corner of the Brillouin zone. 
The parameters of the continuum 
Hamiltonian (near the QBCP) of Eq. \eqref{eq:Hamiltonian} 
are $t_I=(t'+t'')/2$, $t_x=t/2$, $t_z=(t'-t'')/2$. The order parameters are
$Q_1=\frac{1}{4}\sum_\delta \langle c^\dag_{A,i}c_{A,i} -c^\dag_{B,i+\delta}c_{B,i+\delta}\rangle$ (``site nematic''), 
$Q_2= \frac{1}{2}\sum_\delta D_\delta \Re\langle c^\dag_{A,i} c_{B,i+\delta}\rangle$ (``bond nematic''), and 
$\Phi=\frac{1}{2}\sum_\delta D_\delta \Im\langle c^\dag_{A,i} c_{B,i+\delta}\rangle$ (QAH),
where ${\delta}=\pm\hat x/2 \pm \hat y/2$ connects  nearest neighbors. 
$D_\delta=\pm 1$, 
$D_{\pm(\hat x/2+\hat y/2)}=1$ and 
$D_{\pm(\hat x/2-\hat y/2)}=-1$. 
\begin{figure}[hbt]
\begin{center}
\includegraphics[width=0.40\textwidth,trim = 5mm 8mm 5mm 7mm]{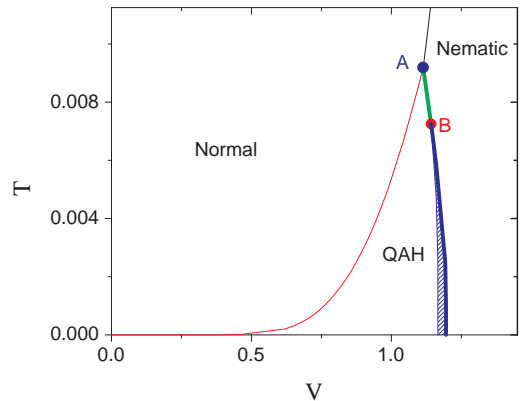}
\end{center}
\caption{(Color online) Mean-field $T-V$ phase diagram of a half-filled checkerboard lattice with $t'/t=0.5$, $t''/t=-0.2$. Nematic and QAH orders coexist in the shaded area. Thick (thin) lines are first (second)
order transitions. $A$ is a bicritical point and $B$ is a critical-end point.}
\label{fig:phase}
\end{figure}

A mean-field theory analysis \cite{Sun-to-come}  (similar to Eq.\eqref{eq:MF}) of the lattice model, 
Eq.\eqref{eq:lattice_model}, yields the $T-V$ phase diagram of Fig. \ref{fig:phase}. 
(The details depend on $t'/t$ and $t''/t$.)
A QAH phase is  found for $V$ small and below a critical temperature. This phase 
has a zero-field quantized Hall conductivity $e^2/h$, and that the quasiparticle spectrum has  
topologically protected chiral edge
states, as predicted by general considerations \cite{qi-2006}.
A site-nematic phase is  found for $V \sim |t'-t''|$, while the bond-nematic is not favored. For $|t'|\ge|t''|$
and $|t''|/|t'| \ll 1$, there is a direct nematic-QAH first order transition. If $|t''|/|t'| \sim 1$, 
there is also a coexisting QAH+nematic phase. 
For other values, there are a direct first-order transition and a coexisting phase.

Near a QBCP, a next-nearest-neighbor attraction $V_{nnn}<0$ \cite{phonons} generates the (irrelevant) 
interaction $U$. 
For $|V_{nnn}/V|\gtrsim 1$, a site-nematic phase ($Q_1$) becomes stable, even at weak coupling. 
A site-nematic phase is stabilized at large enough $V>0$ ($V_{nnn}=0$). 
In the strong coupling limit, at this filling, this system is known to be in 
an insulating site-nematic phase \cite{Kivelson2004}.

{\it Spin-$1/2$ fermions}---
Let us consider briefly the case of a system of spin-$1/2$ fermions at a QBCP.  Details  will be given in
Ref. \cite{Sun-to-come}. 
We consider the four shortest-range interactions on a lattice: 
a) an on-site repulsive Hubbard $U$, 
b) a nearest neighbor repulsion $V$, 
c) a nearest neighbor exchange interaction $J$, 
and d) a pair-hopping term $W$. 
In addition to spin-singlet order parameters ({\it c.f.\/} Eq \eqref{eq:order_parameters}), 
there are  also spin-triplet  order parameters:
\begin{eqnarray}
&&\!\!\!\!\!\!\!\!\!\vec{Q}^t_1=
\avg{\boldsymbol{\Psi}^\dagger(\mathbf{r})(\vec{\tau}\otimes\sigma_z) \boldsymbol{\Psi}(\mathbf{r})},
\;\vec{Q}^t_2=\avg{\boldsymbol{\Psi}^\dagger(\mathbf{r})(\vec{\tau}\otimes\sigma_x) \boldsymbol{\Psi}(\mathbf{r})},
\nonumber\\
&&\!\!\!\!\!\!\!\!\!\vec{S}=
\avg{\boldsymbol{\Psi}^\dagger(\mathbf{r})( \vec{\tau}\otimes I) \boldsymbol{\Psi}(\mathbf{r})},
\; \vec{\Phi}^t=\avg{\Psi^\dagger(\mathbf{r})(\vec{\tau}\otimes\sigma_y) \boldsymbol{\Psi}(\mathbf{r})}.
\label{eq:order-parameters-spin1/2}
\end{eqnarray}
where $\vec{\tau}$ are the three Pauli matrices.
Here, $\vec{S}$ is the spin density. For $\vec{Q}^t_{1}\neq 0$ the QBCP splits into four 
Dirac points displaced along the main axes.
This state has reversed spin polarization
along $x$ and $y$ axes. For a QBCP with $C_4$ symmetry, the charge sector is still $C_4$ invariant, but the 
spin sector
changes sign under a rotation by $\pi/2$.
Thus, $\vec{Q}^t_{1}\neq 0$ is a NSN state \cite{Oganesyan2001,Kivelson2003-short,Wu2007a}. 
$\vec{Q}^t_2\neq 0$ describes NSN order along the diagonals.

A state with  $\vec{\Phi}^t \neq 0$ is a QSH phase~\cite{Kane2005,Bernevig2006,Konig2007,Raghu2008} 
with helical 
edge states~\cite{Wu2006,Xu2006}. In this phase, the two spin components have opposite Hall conductivity.
The QSH topological insulator has a quantized spin Hall conductivity,  and it is a spin triplet version of the QAH phase. 
It is an insulating analog of the  $\beta$-phase of Refs. \cite{Wu2004,Wu2007a}.

A scaling analysis similar to the spinless case  finds six marginally relevant operators, associated
with the four nematic order parameters (singlet and triplet) and the QAH and QSH orders. 
We have obtained the phase diagram in mean-field theory for spin-$1/2$ fermions on a checkerboard lattice 
at a QBCP with $U>V>0$ at $T=0$\cite{thermal}. For simplicity we take $t$, $t'$ and $t''$ such that $t_x \sim t_z$, and $W=0$. Once again all the
susceptibilities are logarithmically divergent with $\chi_{\vec \Phi}=\chi_\Phi$, and $\chi_{\vec Q_i}=\chi_{Q_i}$
($i=1,2$), so the interactions are marginally relevant. For $U>2V>0$, the system
is in the NSN phase at low temperatures. For $2V>U>0$ we find the QAH phase for $J>0$, and the QSH phase for
$J<0$. Both gapped phases are topological insulators. The gaps and
critical temperatures obey a scaling law similar to the spinless case \cite{Sun-to-come}. 
For $U\rightarrow \infty$ (and $J=0$),
there is a 
NSN state, while $V\rightarrow \infty$ 
stabilizes a nematic phase.

Using RG methods and mean-field theory we showed that a system of 
interacting fermions, with or without spin,
at a QBCP 
have topological insulating QAH or QSH phases, 
at arbitrarily weak  
short-range repulsive interactions.   
These perturbatively accessible topological insulating phases are  due to 
spontaneous symmetry breaking, described by order parameters, 
and are not due to spin orbit effects in the band structure. 
At intermediate coupling we also find nematic (and coexisting) phases. Using 
large $N$ methods and a $2+\epsilon$
expansion, we infer the existence of these phases in 3D (similar to those of Ref.\cite{Zhang2009}), 
but at a finite critical coupling\cite{Sun-to-come}. 

We thank S. Raghu and S. C. Zhang for comments. 
This work was supported in part by the National Science Foundation 
under grant DMR 0758462 (EF), and the Office of Science, U.S. 
Department of Energy under Contracts DE-FG02-91ER45439 of the Frederick
Seitz Materials Research Laboratory at the University of Illinois (EF,KS), and DE-FG02-06ER46287 
at the Gaballe Laboratory of
Advanced Materials of Stanford University (SAK,HY), and from SGF at Stanford University (HY).


\begin{thebibliography}{33}
\bibitem[{ban()}]{band-anisotropy}
\bibinfo{note}{$t_x/t_z$ does not flow under the RG to two loop order.}
\bibitem[{pho()}]{phonons}
\bibinfo{note}{$V_{nnn}<0$ may be due to a coupling to phonons.}
\end{thebibliography}

\end{document}